\documentclass[%
reprint,
amsmath,amssymb,
pra,
floatfix,
]{revtex4-2}

\usepackage[english]{babel}
\usepackage{letltxmacro}

\LetLtxMacro{\ORIGselectlanguage}{\selectlanguage}
\makeatletter
\DeclareRobustCommand{\selectlanguage}[1]{%
  \@ifundefined{alias@\string#1}
    {\ORIGselectlanguage{#1}}
    {\begingroup\edef\x{\endgroup
       \noexpand\ORIGselectlanguage{\@nameuse{alias@#1}}}\x}%
}
\newcommand{\definelanguagealias}[2]{%
  \@namedef{alias@#1}{#2}%
}
\makeatother

\definelanguagealias{en}{english}

\usepackage{graphicx}% Include figure files
\usepackage{svg, float}
\usepackage{dcolumn}% Align table columns on decimal point
\usepackage{bm}% bold math
\let\vec\mathbf
\usepackage{hyperref}% add hypertext capabilities
\usepackage{amsmath}
\usepackage{xcolor}
\hypersetup{
    colorlinks,
    linkcolor={red!80!black},
    citecolor={blue!80!black},
    urlcolor={purple!80!black}
}

\usepackage{orcidlink}

\begin{document}

\title{SuperVortexNet: Reconstructing Superfluid Vortex Filaments Using
Deep Learning}

\author{Nick A. Keepfer\,\orcidlink{0000-0002-7803-3713}}
\affiliation{Joint Quantum Centre (JQC) Durham--Newcastle, School of Mathematics, Statistics and Physics, Newcastle University, Newcastle upon Tyne, NE1 7RU, UK}
\author{Thomas A. Flynn\,\orcidlink{0000-0001-9800-594X}}
\affiliation{Joint Quantum Centre (JQC) Durham--Newcastle, School of Mathematics, Statistics and Physics, Newcastle University, Newcastle upon Tyne, NE1 7RU, UK}
\author{Nick G. Parker\,\orcidlink{0000-0002-7238-8492}}
\affiliation{Joint Quantum Centre (JQC) Durham--Newcastle, School of Mathematics, Statistics and Physics, Newcastle University, Newcastle upon Tyne, NE1 7RU, UK}
\author{Thomas P. Billam\,\orcidlink{0000-0003-2456-6204}}
\affiliation{Joint Quantum Centre (JQC) Durham--Newcastle, School of Mathematics, Statistics and Physics, Newcastle University, Newcastle upon Tyne, NE1 7RU, UK}

\date{\today}

\begin{abstract}
We introduce a novel approach to the three-dimensional reconstruction of superfluid vortex filaments using deep convolutional neural networks. Superfluid vortices, quantum mechanical phenomena of immense scientific interest, are challenging to image due to their small dimensions and intricate topology. Here, we propose a deep-learning methodology that serves as a proof-of-principle for fully reconstructing the topology of superfluid vortex filaments.
We have trained a convolutional neural network on a large dataset of simulated superfluid density images obtained by solving the Gross--Pitaevskii equation at scale, enabling it to learn the complex patterns and features inherent to superfluid vortex filaments. The network ingests the integrated density along the axial, coronal, and sagittal directions and outputs the reconstructed superfluid vortex filaments in three dimensions. We demonstrate the success of this approach over a range of vortex densities of simulated isotropic quantum turbulence, enabling access to the characteristic scaling law of the decaying vortex line length.
\end{abstract}

\maketitle

%\section{Introduction}
Superfluid vortices are the fundamental entities of rotation in a frictionless superfluid, consisting of individual thin filaments around which the superfluid rotates either clockwise or anticlockwise with quantised circulation~\cite{Onsager1953, Feynman1955}. The turbulent dynamics of tangled superfluid vortex filaments --- so-called quantum turbulence --- has long been an object of study in liquid Helium~\cite{Donnelly1986, Barenghi2001}. However, the difficult problem of directly imaging the superfluid vortex filaments themselves, which have a vortex core size on the order of Angstroms, was only relatively recently solved using solid tracer particles suspended in the liquid~\cite{Guo2014}.

The advent of ultracold atomic Bose--Einstein condensates has provided another, highly configurable platform for the realisation of superfluid vortex filaments and quantum turbulence~\cite{Wilson2013, Tsatsos2016}. In these systems the vortex core size is typically on the order of microns, and superfluid vortex filaments can be imaged directly through absorption imaging~\cite{Freilich2010, Lamporesi2013, Serafini2017} or other techniques~\cite{Wilson2015}. In quasi-two-dimensional setups, where the vortex filaments are constrained to approximate straight lines along the tightly confined direction~\cite{Rooney2011}, this has enabled impressive realisations of phenomena in so-called two-dimensional quantum turbulence. These include a superfluid von K\'arm\'an vortex street~\cite{Kwon2016} and Onsager vortex clustering~\cite{Gauthier2019, Johnstone2019}. In three dimensions, absorption imaging of outcoupled atoms has been used to observe the real-time dynamics of straight vortex filaments~\cite{Freilich2010}, and interacting, curved, vortex filaments in an elongated condensate~\cite{Serafini2017}. The ability to image the tangle of superfluid vortex filaments in three-dimensional quantum turbulence in similar detail is highly desirable in order to validate theoretical predictions. For example, the tangle is predicted to have either a classical or a so-called ultraquantum form, depending on how the tangle is generated, with different scaling laws for the decay of the vortex line length in each case~\cite{stagg_ultraquantum_2016}. However, with standard imaging techniques for condensates yielding images projected onto a particular plane, direct observation of a three-dimensional tangle of superfluid vortex filaments in these systems remains elusive.

In recent years, advances in machine learning, and more specifically
convolutional neural networks (CNNs), have opened up new possibilities for
imaging and analysing complex systems \cite{lecun_deep_2015}. CNNs have
demonstrated remarkable success in image recognition tasks
\cite{alzubaidi_review_2021} and were recently applied to detection of superfluid vortices in two dimensions for both synthetic \cite{metz_deep-learning-based_2021} and experimental~\cite{kim_vortex_2023} data. They are therefore a promising tool for the more challenging case of detecting superfluid vortex filaments in three dimensions. Filament-like structures are ubiquitous in the natural
world. In the specific use-case of identifying filament-like features, CNN's
have found success in applications such as identifying road-cracks
\cite{zhang_road_2016,di_benedetto_u-net-based_2023}, retinal vessel
segmentation \cite{hajabdollahi_low_2018,chala_automatic_2021}, root structure
analysis in agriculture \cite{atanbori_convolutional_2019}, river and stream
analysis form satellite data \cite{ling_measuring_2019} and even cosmological
filament detection \cite{gheller_deep_2018,aragon-calvo_classifying_2019}.

In this paper we introduce SuperVortexNet (SVN), a neural network designed to process
two-dimensional integrated density profiles along the axial, saggital and
coronal directions into a three-dimensional binary voxelised array that
indicates the presence or absence of a superfluid vortex filament. 
We have trained a variant of SVN (SVN-B) which is robust to noise in its input integrated density profiles. We demonstrate on simulations of isotropic quantum turbulence that this recovers the three-dimensional vortex distribution over a range of vortex densities.  This presents a breakthrough towards experimental imaging of three-dimensional quantum turbulence and opens the door to 
combine SVN with experimental two-dimensional absorption images so as to observe real three-dimensional superfluid vortex tangles in detail.

\section*{Results}

\begin{figure*}[ht]
\begin{center}
    %\includesvg[width=30pt]{pipeline.svg}
    \includegraphics[width=1.99\columnwidth]{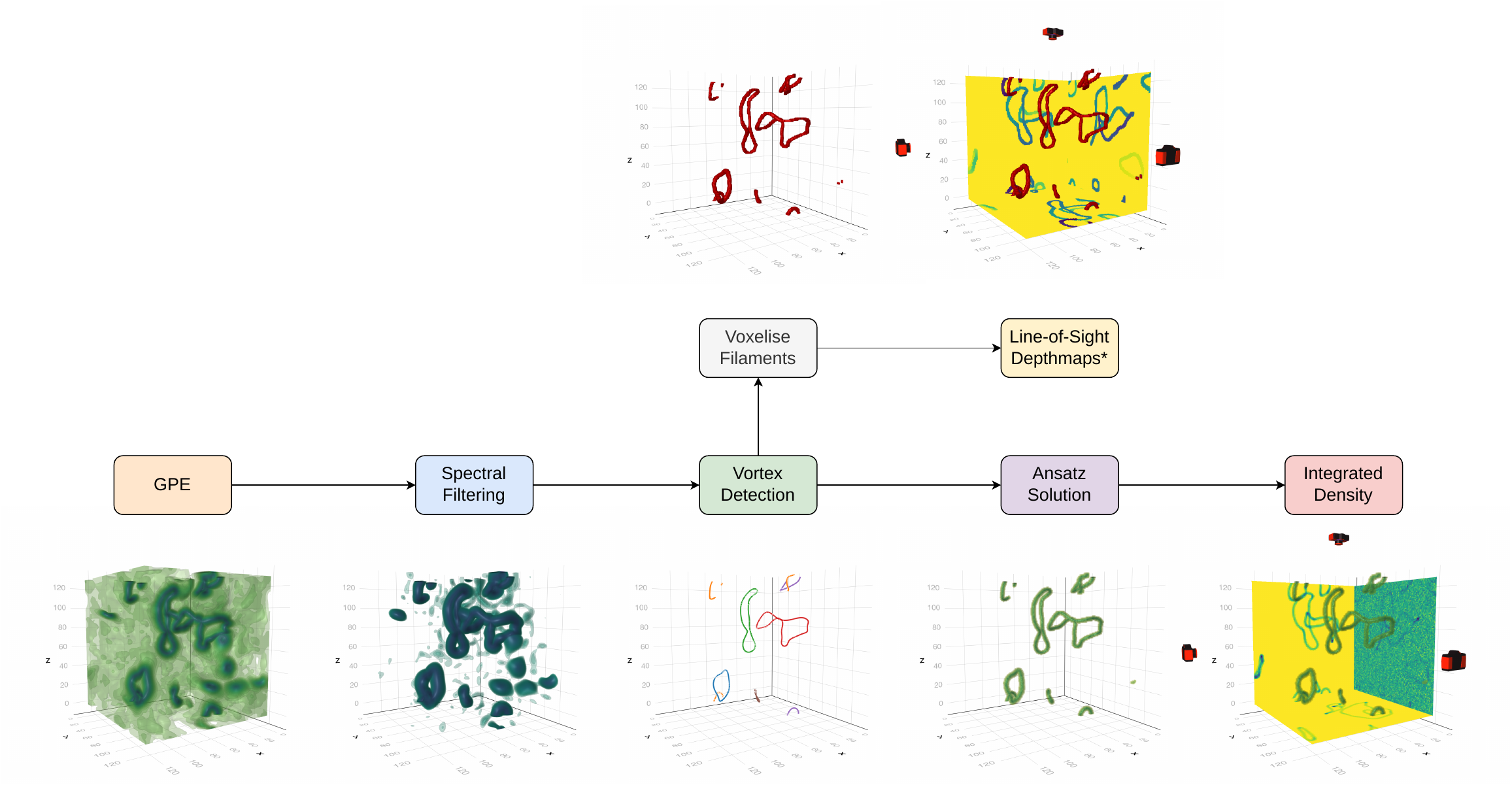}
\end{center}
\caption{Data collection pipeline: Initially, we run a %weakly-dissipative 
GPE simulation of decaying ultraquantum turbulence, similar to Ref.~\cite{berloff2002,stagg_ultraquantum_2016}. The numerical wavefunctions then pass through a spectral filter to remove high-frequency noise. Following this, we run a vortex detection algorithm using a plaquette-based method \cite{Foster_Vortex_2010}. We then use the detected vortex filaments to construct a density field using an ansatz solution \cite{fetter_rotating_2009} for the vortex cores; we refer to this as a `cleaned' density field with this specific, realistic configuration of vortex filaments but no noisy density fluctuations. From this density field we can obtain the integrated density profiles which serve as the inputs (images) to our model. We also generate the ground truth (labels) of vortex filaments in a voxelised grid by using the detected vortex filament positions to populate a binary discretised grid and performing morphological inflation (see ``Methods''). 
As an additional aid to train our model, we use the voxelised filaments to create line-of-sight depthmaps. In these the distance to each vortex is encoded as a value up to unity along the axial, saggital, and coronal directions. $^*$We stress here that the depthmaps are only used as a training aid, and are not needed at inference time for the model.}
\label{fig:data_pipeline}
\end{figure*}

SuperVortexNet consists of a modified U-Net~\cite{ronneberger_u-net_2015} followed by the expanding path of a custom V-Net~\cite{milletari_v-net_2016}, constructed and trained using Flux.jl \cite{innes_flux_2018,innes_fashionable_2018} (see ``Methods'' for details of the motivation of this deep learning approach, the model architecture and training, and the loss functions used). Our data collection pipeline, shown in Fig.~\ref{fig:data_pipeline}, provides a point of reference from which we outline the operation of the network. We gathered a training dataset of three-dimensional isotropic quantum turbulence from simulations of the weakly-dissipative Gross--Pitaevskii equation (GPE) simulations starting from highly non-equilibrium initial conditions.  This previously studied approach \cite{berloff2002,stagg_ultraquantum_2016} is known to simulate a thermal quench of the system and give rise to an isotropic tangle of vortices which decays over time (see ``Methods'' for the details of training data acquisition). Post-processing of the simulation data yields three datasets: a ground truth, in the form of detected superfluid vortex filaments; input data, in the form of cleaned integrated density along the axial, coronal and saggital directions, which are embedded into the three colour channels of an input image; and additional data, used only as a training aid and not for inference, in the form of line-of-sight depthmaps along each direction. We trained two variants of the network. Variant SVN-A learned directly from cleaned integrated densities, whereas variant SVN-B learned from integrated densities with added noise. Specifically, we added independent Gaussian random noise to each integrated density perspective with variance $\sigma=0.1$, corresponding to $10$\% of the background bulk density. This added noise can obscure or distort the vortex signatures, making it more difficult for the models to accurately identify and localise vortex filaments. This provides a valuable test of the model's capability to maintain high accuracy and reliability in the presence of disturbances that are representative of practical experimental conditions.

\begin{figure}[t]
    \centering
    \includegraphics[width=0.5\textwidth]{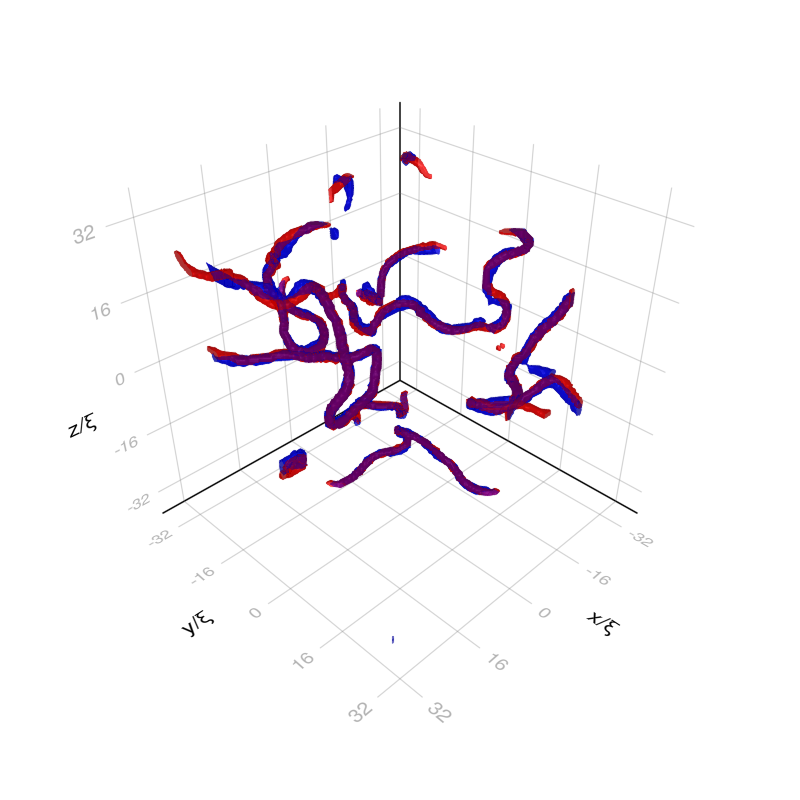}
    \caption{Ground truth (red) and SVN-A prediction (blue) for unseen test data. Both are represented as 3D meshes. The ground truth mesh is simply a voxelisation of vortex core positions whilst the prediction mesh is generated by applying a probability threshold of $\epsilon=0.5$ to the 3D probability field outputted by the neural network. This threshold is used to determine the presence of a vortex within each voxel of the prediction volume. Voxels with a probability equal to or greater than $\epsilon$ are considered to contain a vortex.}
    \label{fig:prediction}
\end{figure}

Our primary result for SVN-A is shown in Fig.~\ref{fig:prediction}. In the absence of density noise SVN-A is able to reconstruct complicated three-dimensional superfluid vortex filaments from unseen integrated-density test data with qualitatively impressive performance. Quantitatively, when evaluating the network on our test set, we see an average Dice loss coefficient of 0.244 and an average Intersection over Union (IoU) of 0.538.

Going beyond individual filament configurations, we tested SVN-B's ability to extract theoretically important measures of quantum turbulence from noisy data by computing the line length of the reconstructed vortex filaments on unseen data from simulations of decaying quantum turbulence. We generated an additional, unseen test dataset for this purpose using the same simulation method as outlined above for our training and test data, but run without dissipation and over a far longer time period than the simulations used to train the model. This was repeated over an ensemble of five trajectories with randomised initial conditions. Note that the particular state of quantum turbulence produced, the so-called ultraquantum turbulence, is characterised by the vortex line length $L$ decaying according to the scaling law $L \sim t^{-1}$\cite{stagg_ultraquantum_2016}.

\begin{figure}
    \centering
    \includegraphics[width=\linewidth]{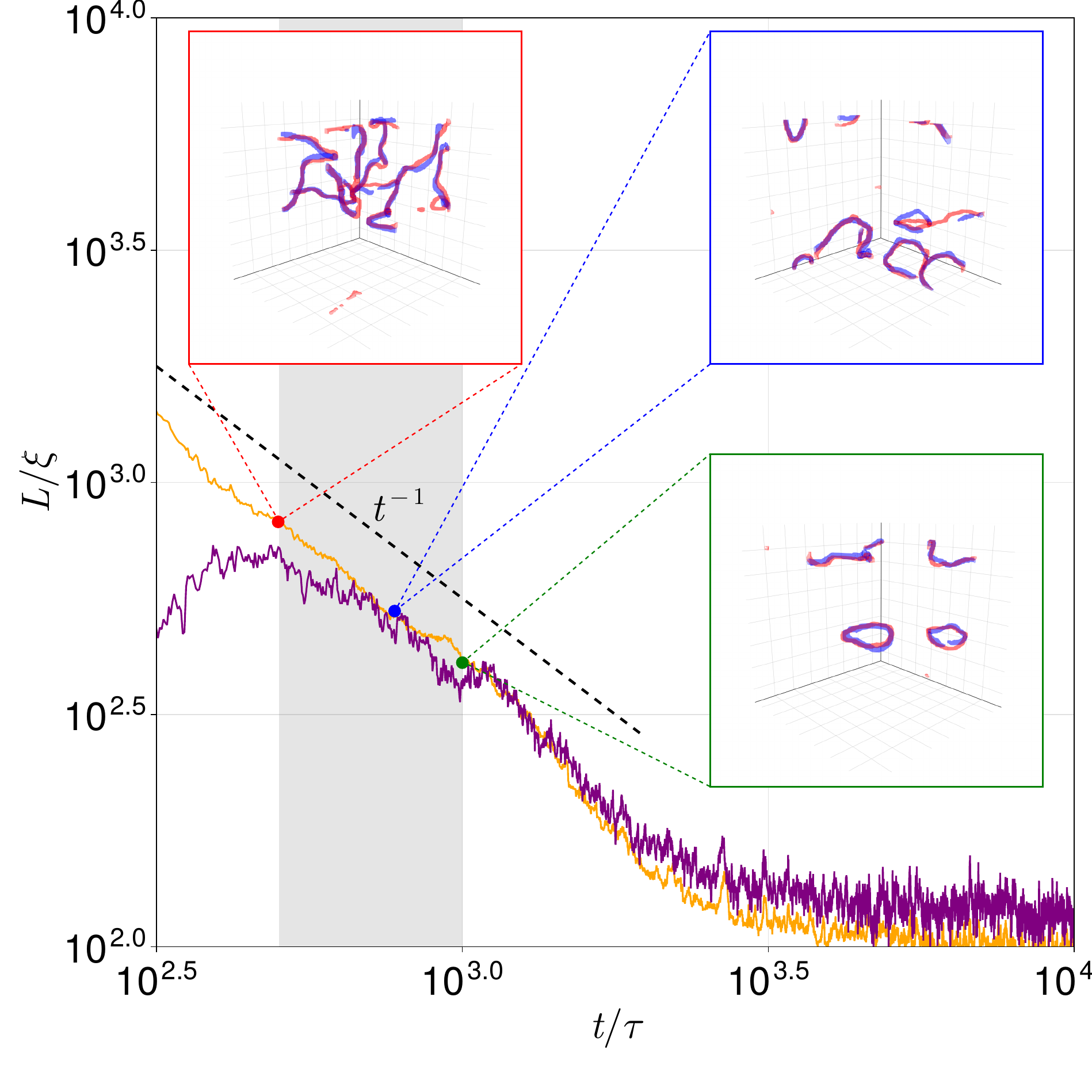}
    \caption{Time evolution of the averaged vortex line length $L$ for an ensemble of five simulations of decaying quantum turbulence. The ground truth and SVN-B prediction for unseen, noisy, test data are depicted as orange and purple lines, respectively. A dashed black line, scaling with $t^{-1}$, illustrates the characteristic decay law of ultraquantum turbulence \cite{stagg_ultraquantum_2016}. Insets show 3D representations of vortex configurations at specific times $t=500\tau$ (red), $t=775\tau$ (blue) and $t=1000\tau$ (green). In each inset, the true vortex configuration is shown in red and the prediction in blue. A shaded grey region indicates the time window over which the neural network received training data.}
    \label{fig:linedecay}
\end{figure}

The extraction of proxy measurements such as this using SVN provides an innovative way to quantify and analyse the complex topology of superfluid vortices. Computing the total vortex line length in the system from the SVN reconstruction simply involves tallying the lengths of all voxelised segments that make up the reconstructed vortex filaments, taking the ablation of the vortices into account using morphological erosion (see ``Methods''). To validate the accuracy of our vortex line length estimation, we compared the estimated vortex line lengths from SVN-B's reconstructions with the actual lengths calculated directly from the GPE simulations in our additional test dataset, as depicted in Fig.~\ref{fig:linedecay}. The close correspondence observed in Fig.~\ref{fig:linedecay}  underscores the reliability of SVN in capturing this critical aspect of quantum turbulence. Despite SVN-B being trained on realisations from $500\tau$ to $1000\tau$ as indicated by the shaded region, we achieve good accuracy beyond this timescale. That suggests that our model is successfully generalising to different data outside of the original test set, which in turn suggests that SVN has become proficient in the task of extracting the topology of vortices just from the integrated density profiles alone. At times before $500\tau$ there is a clear deviation from the ground truth result. However, we expect the very high density of vortex filaments in the tangle prior to this time to provide much occlusion of filament segments by other filament segments along every direction in the tangle. High numbers of occlusions make inference more difficult for the model, and we believe the line length deviation at times before $500\tau$ can be understood on these grounds. It is important to note that the SVN line length prediction clearly recovers the $L\sim t^{-1}$ scaling behaviour to a similar level of accuracy as the GPE simulations themselves, highlighting the potential of the SVN approach to measure and verify decay laws in quantum turbulence experiments.

\section*{Discussion}

In this study, we have introduced a novel deep-learning methodology for the three-dimensional reconstruction of superfluid vortex filaments.
The SVN model was trained on a large dataset of simulated superfluid density images, generated from large-scale GPE simulations of isotropic quantum turbulence.
We find that SVN meets the challenge of reconstructing three dimensional vortex filament structure and line length from limited two-dimensional, integrated-density information from axial, coronal, and sagittal perspectives. Furthermore, by applying SVN to simulations of decaying, isotropic quantum turbulence, we demonstrate that it successfully measures the vortex line length over a range of vortex densities and allows verification of the expected scaling law for the decay of vortex line length.

We have focused on a clear, proof-of-principle demonstration of the methodology in this work. However, we note that the integrated-density information that the present version of SVN ingests is similar to the information that would be experimentally accessible by absorption-imaging outcoupled fractions of a Bose--Einstein condensate along multiple axes~\cite{Freilich2010, Serafini2017}. Hence, this work is a major step towards experimental observation of a three-dimensional vortex tangle and of experimentally measuring the decay of vortex line length, a key probe of the nature of the turbulent state. 

As the methodology shows promise, there are several avenues for future work. Optimising the architecture and hyperparameters of our deep-learning model could lead to further improvements in the efficiency and accuracy of vortex reconstruction. A logical next step would also be the integration of real-world experimental data to validate the efficacy of our approach, although we expect the details of any experimental imaging system to necessitate some custom training of the network. Moreover, given the recent advances in transformer-based \cite{vaswani_attention_2017, dosovitskiy_image_2021} neural networks for a variety of complex tasks, we envisage the adaptation of such architectures as a promising avenue for enhancing the performance and interpretability of the methodology.

\section*{Methods}

\subsection*{Deep Learning Approach}

In the context of deep CNN approaches to identifying filamentary structures the U-Net architecture has proven to be particularly effective, thanks to its unique design that emphasises the
localisation of features within an image \cite{ronneberger_u-net_2015}. The
U-Net architecture consists of a contracting path to capture context and a
symmetric expanding path that enables precise localisation. This design has
made it a preferred choice for tasks that require the detection of intricate
patterns and structures, such as the filament-like features mentioned above.
Similarly, the V-Net architecture has emerged as a powerful tool for volumetric
segmentation, particularly in the context of medical imaging
\cite{milletari_v-net_2016}. The V-Net extends the principles of the U-Net to
3D data, employing a fully convolutional volumetric neural network that excels
in capturing complex spatial relationships. 
As mentioned above, SuperVortexNet consists of a modified U-Net followed by the expanding path of a custom V-Net.
Instead of re-using a
CNN for each individual view, as in multi-view networks
\cite{choy_3d-r2n2_2016,xie_pix2vox_2019}, we instead embed our different views
into the colour channels of our input image. While multi-view networks have
shown proficiency in generating 3D volumes from 2D images of classical objects
with relatively fixed topology, their application to quantum vortices encounters
significant challenges. Unlike typical 3D objects such as chairs or cars, which
offer topological and dimensional regularities for the neural network to
generalise from, quantum vortices are quasi-one-dimensional filaments with a
consistent thickness but non-fixed, fluid topology. This makes it challenging
to establish a `generic' shape or structural basis to generalise across
instances. The inherent limitations in capturing such low-dimensional and
topologically variable features make multi-view networks an unsuitable choice
for accurate 3D reconstruction of quantum vortices. Our approach instead allows
us to efficiently process multiple views within a single CNN, reducing the
complexity and computational requirements of the model and allowing contextual
flow between channels. This integration allows the model to
simultaneously capture information from multiple perspectives, enhancing the
feature extraction process. The model architecture is illustrated in
Fig.~\ref{fig:network_arch} and is constructed and trained using Flux.jl
\cite{innes_flux_2018,innes_fashionable_2018}.

\subsection*{Model Architecture}

The architecture of our network is designed to address the unique challenges of
three-dimensional reconstruction of superfluid vortices, particularly focusing
on the intricate filament-like structures inherent to these phenomena.
\begin{figure*}
\begin{center}
\includegraphics[width=1.9\columnwidth]{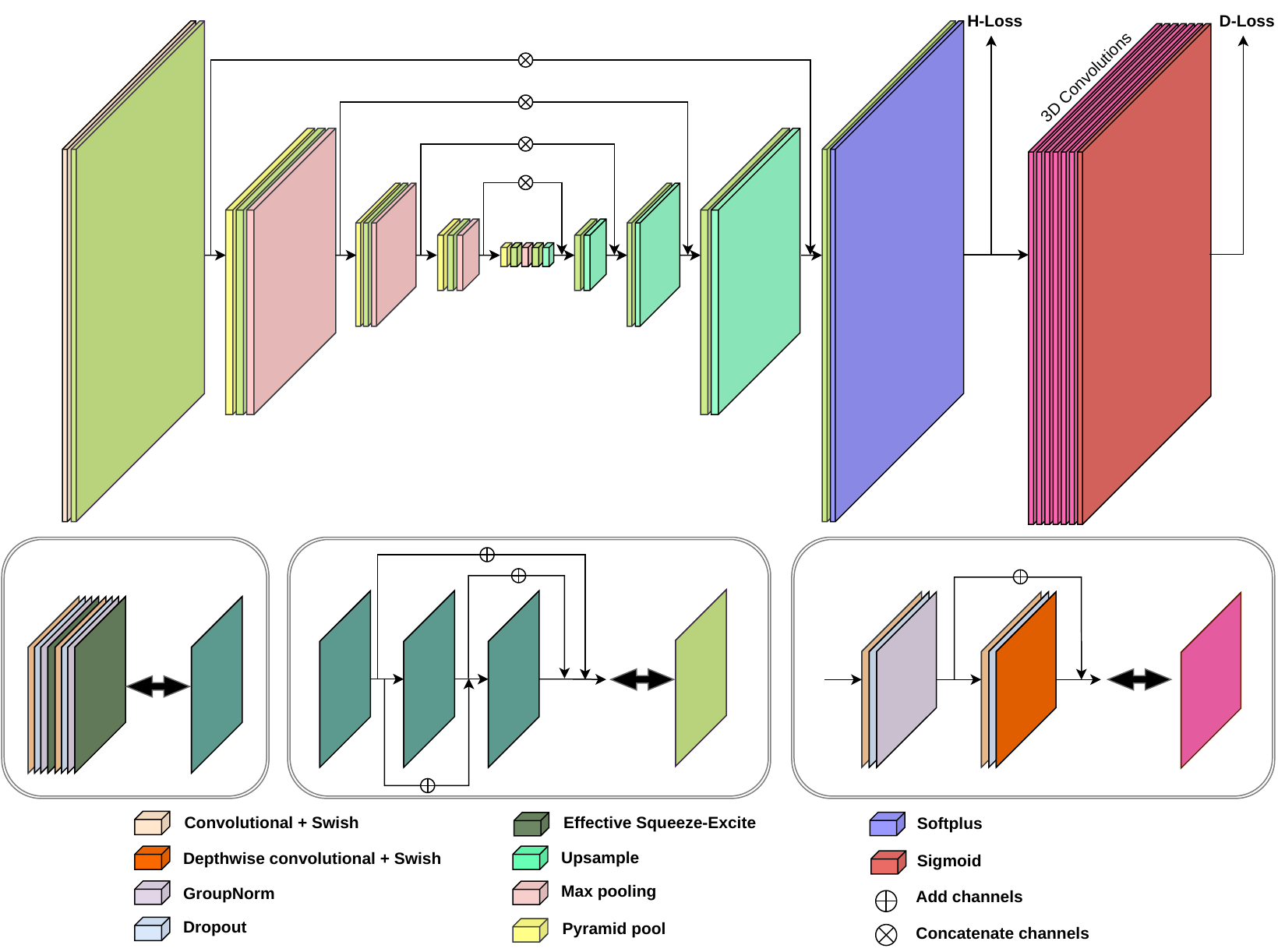}
\end{center}
\caption{Schematic representation of the deep convolutional neural network
designed for the three-dimensional reconstruction of superfluid vortices. Key
modules include convolutional layers with Swish activation, depthwise
convolution, GroupNorm regularisation, and an effective Squeeze-Excite
mechanism. The architecture also integrates max pooling and pyramid pooling modules to
capture spatial hierarchies. H-Loss and D-Loss correspond to the Huber loss and
Dice loss metrics respectively. The bottom insets detail specific functional
blocks and their interactions, while the legend provides a clear mapping of
colours to individual operations or modules.}
\label{fig:network_arch}
\end{figure*}
In Fig.~\ref{fig:network_arch} we detail the key components and innovative aspects of our architecture. The initial part of the network, modelled after a U-Net structure, consists of several 2D convolutional layers designed to map integrated density profiles into line-of-sight depthmaps for each of the axial, saggital and coronal planes simultaneously. The operational cycle of our network is systematically structured, beginning with a convolution layer that extracts feature maps from the input data. Following this, the ``swish'' \cite{ramachandran_searching_2017} activation function introduces non-linearity, allowing the network to learn complex patterns. To prevent over-fitting and enhance generalisation, a Dropout \cite{srivastava2014dropout} layer randomly deactivates a subset of neurons during training. Subsequent normalisation, specifically group normalisation \cite{wu_group_2018}, then stabilises the learning process by normalising the input features. Finally, we utilise Effective Squeeze-Excitation blocks, as described in \cite{lee_centermask_2020}, to allow the network to re-calibrate the channel-wise feature responses, fine-tuning the network's focus on relevant features. This operational cycle is present throughout the entirety of the network. In addition, both ResNet-style Skip-Connections \cite{he_deep_2015} and element-wise addition Skip-Connections are employed, serving to mitigate the vanishing gradient problem by improving gradient flow, thereby enhancing the training stability and model performance. 

The contracting path of the U-Net utilises Pyramid-Pooling modules \cite{lazebnik_beyond_2006,he_spatial_2015,zhao_pyramid_2017} which help to encode contextual spatial information into feature maps at different scales. Specifically, for a feature map with dimensions \texttt{[height, width, channels]}, we employ pooling operations with kernels corresponding to powers of $2$, up to $2^{N_s-1}$, where $N_s$ denotes the height or width of the feature map at that specific depth in the network. The MaxPool operation \cite{scherer_evaluation_2010} is also utilised at this point in our network to reduce the spatial dimensions of the feature maps. This process, by selecting the maximum value over a specified window size, effectively condenses the feature information. It not only helps in reducing computational load and controlling overfitting but also aids in retaining the most significant features within each pooling region. In contrast, the expanding path of the U-Net features upsampling operations which perform bilinear interpolation on the feature maps to upscale the dimensionality to match the corresponding image size on the contracting path. At the final layer of the U-Net we apply a Softplus \cite{zheng_improving_2015} activation function to ensure positive outputs.

The V-Net section of our network focuses on converting the generated line-of-sight depthmaps into a 3D voxel grid that represents the existence or absence of vortex lines. This part of the architecture is made up of 3D upsampling deconvolutional layers \cite{zeiler_deconvolutional_2010}.  At each upsampling block, we add a skip-connection path and an additional depth-wise convolutional operational block, which helps in gradient flow, preserving the spatial hierarchies and feature learning capabilities of the network respectively. At the final layer of SVN we apply a Sigmoid activation function to map the output to a probability distribution signifying the likelihood of a vortex being present at a particular voxel position.

\subsection*{Loss functions}

We utilise a multi-task learning framework for our architecture, where our main components, the U-Net and V-Net expanding path, are trained concurrently but optimised for distinct objectives. For the first portion of the network we aim to convert integrated density profiles into depthmaps, making use of the Huber loss function \cite{huber_robust_1964}. The Huber loss provides a balanced compromise between the Mean Squared Error and the Mean Absolute Error, effectively handling outliers without suffering from large gradients that could potentially destabilise the training process. The Huber loss is formulated as:
\begin{equation}
\mathcal{L}_{\text{Huber}}(\theta) = \sum_{\text{batch}, c, x, y} \phi(P_{c,x,y} - G_{c,x,y})\,.
\end{equation}
Here, \( \phi(z) \) represents the Huber function and is defined as:
\[
\phi(z) = 
\begin{cases} 
\frac{1}{2} z^2 & \text{if } |z| \leq \delta \\
\delta (|z| - \frac{\delta}{2}) & \text{otherwise.}
\end{cases}
\]
with $\delta$ as a hyperparameter we set to unity. The summation $\sum_{\text{batch}, c, x, y}$ covers all batches, channels, and pixels. $P$ and $G$ signify the predicted and ground truth values, respectively, for each channel $c$ at pixel $(x, y)$. We use the calculated depthmaps to serve as training labels for this loss function but stress that these depthmaps are only neccessary during training and not at inference.
%A softplus activation function \cite{glorot_deep_2011} is employed in the final layer of the U-Net to ensure strictly positive output values.
The second part of the network, tasked with projecting the depthmaps into a 3D voxelised grid, utilises the Dice loss function \cite{milletari_v-net_2016}, defined as:
\begin{equation}
\mathcal{L}_{\text{Dice}}(\theta) = 1 - \frac{2 \sum_{\text{batch}, x, y, z} P_{x,y,z} \cdot G_{x,y,z}}{\sum_{\text{batch}, x, y, z} P_{x,y,z}^2 + G_{x,y,z}^2}.
\label{eq:dice}
\end{equation}
This loss function is particularly effective for segmentation tasks involving imbalanced classes, as it ensures an equitable representation of both the foreground and background within the 3D grid. The combined use of Huber and Dice loss functions at different points of the network, each tailored to its respective task, contributes to the overall robustness and accuracy of our model. Therefore, the complete loss function for SVN follows as
\begin{equation}
    \mathcal{L} = \mathcal{L}_{\text{Huber}} + \lambda\mathcal{L}_{\text{Dice}}\,,
    \label{eq:loss}
\end{equation}
where we have introduced the hyperparameter $\lambda=0.01$ to balance the loss contributions.

\subsection*{Training Data Acquisition}

Neural networks require vast amounts of data to effectively train and generalise to new, unseen data. The quality and diversity of the training dataset significantly influences the performance of the model. In this section we introduce our methodology for gathering a suitable dataset for SVN.

We begin by modelling the dynamics of a superfluid system using the Gross--Pitaevskii equation (GPE) \cite{pethick_boseeinstein_2008,pitaevskii_bose-einstein_2016,barenghi_primer_2016}. Using this, we can parameterise a weakly-interacting superfluid Bose-Einstein condensate via a mean-field complex wavefunction, $\Psi(\vec{r},t)$, governed by the equation
\begin{equation}
   i \hbar \frac{\partial \Psi}{\partial t}= \left(1-i\gamma\right)\left[-\frac{\hbar^{2}}{2 m}\nabla^2 + V(\vec{r}, t)+g|\Psi|^{2}\right] \Psi. 
\end{equation}
Here we have introduced the reduced Planck constant $\hbar$, the interatomic interaction strength $g$ to model binary inelastic collisions, and the atomic mass $m$. We also introduce a weak dimensionless dissipation parameter $\gamma=0.01$ to accelerate the dynamics of the system (although we set $\gamma=0$ when generating the additional test dataset for the decay of ultraquantum turbulence). Limiting our scope to the case of repulsive inter-atomic interactions, we assign $g>0$ as is typical in experiments. The atomic density follows as
$n(\vec{r},t) = \lvert \Psi(\vec{r},t) \rvert^2$. We then characterise the natural length scale, or healing length, as $\xi = \hbar/\sqrt{m\mu}$ and the natural timescale as $\tau = \hbar/\mu$, where $\mu$ is the chemical potential, where time-independent solutions of the GPE satisfy $i \hbar \Psi_{t}=\mu \Psi$. We may then write the natural units of density and corresponding speed of sound as $n_0 = \mu/g$, and $c = \sqrt{n_0g/m}$ respectively.

We now detail our data acquisition pipeline, which is depicted graphically in Fig.~\ref{fig:data_pipeline}. First, we initialise our GPE simulation with a highly non-equilibrium state as described in \cite{stagg_ultraquantum_2016} to emulate a thermal quench. We then solve the GPE on a $128\times128\times128$ homogeneous periodic grid and run for $1000\tau$, taking samples when the vortex tangles are less complicated past $500\tau$ once every $2.5\tau$. We do this $50$ times to generate $10000$ samples and then use standard data augmentation techniques such as random translations and flips to double this amount. The resulting dataset is then randomly split into a training and test set with $18000$ and $2000$ samples, respectively.

The resultant numerical wavefunctions following the quench contain a high amount of noisy density fluctuations. We use a spectral filtering technique to filter out high-frequency noise. We then use a vortex filament detection algorithm based on the plaquette method \cite{Foster_Vortex_2010} to locate the vortex filaments. We generate a new three-dimensional density field based on the identified structures using an analytic ansatz for the vortex core profile \cite{fetter_rotating_2009}; this produces a `cleaned' density field that contains all of the vortex filament information present in the original state, without the density noise. Previous studies have demonstrated resiliency for CNNs against noise; however in the context of CNNs for quantum vortex detection \cite{metz_deep-learning-based_2021} we expect each magnitude of noise requires a full retraining of the network, which would be costly for our comparatively more complex network. From the `cleaned' density field we take integrated density profiles from the axial, coronal, and saggital directions. These clean profiles are used directly as inputs to the SVN-A variant of the model, and are used after adding Gaussian random noise with variance $\sigma=0.1$ as inputs to the SVN-B variant of the model. Our label data is generated by converting the vortex positions into a voxelised array, wherein we map the continuous points that form the core of each vortex onto the corresponding voxels within a uniformly structured grid. This process effectively converts the spatial information of vortex cores into a discrete, grid-based representation suitable for use with SVN.

To address the class imbalance inherent in our data --- where vortices occupy a relatively small proportion of the spatial domain compared to the non-vortex space --- we apply a morphological dilation to each vortex filament. This dilation process involves expanding the voxels that represent the vortex filaments by a small, predefined amount. The effect is similar to `inflating' the filament, thus increasing the number of voxels representing vortex presence. In calculating the line length of the vortex filaments we subsequently utilise the technique of morphological erosion. This process is the inverse of dilation, where voxels at the boundaries of an object are removed, effectively `shrinking' the representation. This mitigates any adverse effects from the dilation process, enabling a more accurate measure of the vortex line length on the voxelised grid.

\subsection*{Training and evaluation}

In training SVN we use a small batch size of 4 due to computational constraints. We use the Adam Optimiser with a learning rate of $\eta=0.001$ and momentum decay rates $\beta_1 = 0.9$, $\beta_2 = 0.999$. Implementing the loss function as defined in Eq.~\ref{eq:loss}, we train SVN-A for approximately 100 epochs, using the early stopping technique to prevent overfitting. After SVN-A has finished training, we utilise transfer learning, using this model as the starting point to train SVN-B, where Gaussian random noise is added to each channel of the input images. To evaluate the performance of the two variants, we consider the average dice loss coefficient, as defined in Eq.~\ref{eq:dice} as well as introducing the Intersection over Union metric
\begin{equation}
\text{IoU} = \frac{\sum_{\text{batch}, x, y, z} I(P_{x,y,z} > \epsilon) \cdot I(G_{x,y,z})}{\sum_{\text{batch}, x, y, z} I[I(P_{x,y,z} > \epsilon) + I(G_{x,y,z})]}\,,
\end{equation}
where $P_{x,y,z}$ denotes the predicted probability at the voxel located at coordinates $(x,y,z)$ in the prediction volume, while $G_{x,y,z}$ represents the corresponding ground truth value at the same voxel location. The threshold $\epsilon$ is a value that determines whether a voxel is classified as containing a vortex or not, based on the predicted probability. The function $I$ is an indicator function that returns $1$ if its argument is true, and $0$ otherwise. We calculate the mean value of these metrics over each element of the validation dataset, as presented in Table~\ref{tab:performance_metrics}. Importantly, we notice here the expected superior performance of SVN-A over SVN-B, due to the added complexity the added noise brings. 
\begin{table}[t]
\centering
\begin{tabular}{|l|c|c|}
\hline
Model Variant & Mean Dice Coefficient & Mean IoU \\ \hline
SVN-A         & 0.2444         & 0.5387 \\
SVN-B         & 0.3983         & 0.3632 \\ \hline
\end{tabular}
\caption{Comparison of Mean Dice Coefficient and IoU with $\epsilon=0.5$ for SVN-A and SVN-B over each sample in the unseen validation dataset.}
\label{tab:performance_metrics}
\end{table}
All simulations, post-processing, visualisation, neural network construction and training were performed using Julia \cite{bezanson_julia_2015,omlin_solving_2021,danisch_makiejl_2021,innes_flux_2018}, and we utilise CUDA~\cite{cudaRef} to offload calculations to a graphics processing unit. 

\section*{Acknowledgments}
The authors acknowledge support from the UK Engineering and Physical Sciences
Research Council (Grants No. EP/T015241/1, and No. EP/T01573X/1). T. A. F.
also acknowledges support from the UK Engineering and Physical Sciences
Research Council (Grant No.  EP/T517914/1). This work made use of the Rocket
High Performance Computing service at Newcastle University. This work made use of the facilities
of the N8 Centre of Excellence in Computationally Intensive Research (N8 CIR)
provided and funded by the N8 research partnership and EPSRC (Grant No.
EP/T022167/1). The Centre is co-ordinated by the Universities of Durham,
Manchester and York.

\newpage
% Changed to manual file to avoid Zotero shenanigans 
\bibliography{refs}% Produces the bibliography via BibTeX.

\end{document}